\documentclass[3p,times,procedia,authoryear]{elsarticle}
\usepackage[utf8]{inputenc}
\newif\ifphp\phptrue
\phpfalse
\ifphp
\usepackage{ecrc}

\volume{00}

\firstpage{1}

\journalname{Physics Procedia}

\runauth{S. Schippers}

\jid{phpro}

\jnltitlelogo{Physics Procedia}

\CopyrightLine{2011}{The Authors. Published by Elsevier B. V.}
\else
\nopreprintlinetrue
\fi

\usepackage{amssymb}

\bibpunct{(}{)}{,}{a}{}{,}

\begin{document}

\begin{frontmatter}



\ifphp
\else
\vspace*{\fill}
\fi

\title{Electron-ion merged-beam experiments\\ at heavy-ion storage rings}


\author{Stefan Schippers\corref{ca}}
\cortext[ca]{\textit{Email address}: Stefan.Schippers@physik.uni-giessen.de\\~}
\address{Institut f\"{u}r Atom- und Molek\"{u}lphysik, Justus-Liebig-Universit\"{a}t Giessen, 35392 Giessen, Germany}
\begin{abstract}
In the past two decades, the electron-ion merged-beams technique has extensively been exploited at heavy-ion storage rings equipped with electron coolers
for spectroscopic studies of highly charged ions as well as for measuring absolute cross sections and rate coefficients for electron-ion recombination and electron-impact ionization of multiply charged atoms ions. Some recent results are highlighted and future perspectives are pointed out, in particular, in view of novel experimental possibilities at the FAIR facility in Darmstadt and at the Cryogenic Storage Ring at the Max-Planck-Institute for Nuclear Physics in Heidelberg.
\end{abstract}


\end{frontmatter}





\section{Introduction}\label{sec:intro}

The electron-ion merged-beams technique has extensively been exploited at heavy-ion storage rings equipped with electron coolers for the study of collisions of free electrons with highly charged ions.  Two main lines of research are pursued, i) the spectroscopy of highly charged ions aiming at precise measurements of relativistic, nuclear and QED effects and ii) the determination of reliable absolute rate coefficients of electron-ion recombination and electron impact ionization for applications is astrophysics and fusion research. These fields of research face a bright future with upcoming new facilities such as the Cryogenic Storage Ring (CSR) \citep{Krantz2011} at the Max-Planck-Institute for Nuclear Physics in Heidelberg, Germany, the TSR at HIE-ISOLDE \citep{Grieser2012} at CERN in Geneva, Switzerland and the Facility for Antiproton and Ion Research (FAIR) \citep{Stoehlker2014} in Darmstadt, Germany. A comprehensive overview over the physics of electron-ion collisions has been given \citet{Mueller2008a}. The purpose of the present paper is to highlight recent results and to indicate future opportunities.

\ifphp\else
\clearpage
\fi

\section{Experimental technique}\label{sec:exp}

\begin{figure}
\centering{\includegraphics[width=0.7\textwidth]{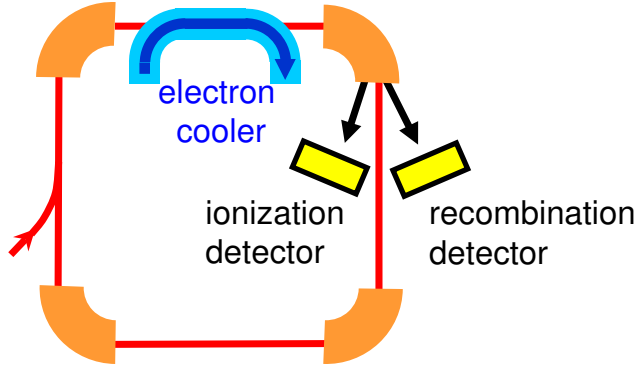}}
\caption{\label{fig:principle} Sketch of the experimental arrangement for electron-ion collision measurements at a heavy-ion storage ring.}
\end{figure}

Detailed experimental studies of, in particular, electron-ion recombination have become feasible by the implementation of the electron-ion merged-beam technique \citep{Phaneuf1999} at heavy-ion storage rings which are equipped with electron coolers \citep[Fig.~\ref{fig:principle},][]{Mueller1997c,Schuch2007a,Schwalm2007}. From an accelerator-physics point of view, electron cooling \citep{Poth1990} is used to reduce the phase space, i.e., the spatial and momentum spread of the stored ion beam. To this end, in one of the straight sections of the storage ring a cold electron beam is magnetically guided collinearly with the ion beam. When electrons and ions have the same average velocity in the laboratory frame, corresponding to zero electron-ion collision energy in the center-of-mass frame, the hot ions efficiently transfer energy to the cold electrons by Coulomb collisions. For the measurement of photorecombination (PR)
\begin{equation}\label{eq:PR}
e^- + A^{q+} \to A^{(q-1)+} +\mathrm{~photons}
\end{equation}
and electron impact ionization (EII)
\begin{equation}\label{eq:EII}
e^-+A^{q+} \to A^{(q+1)+} + 2e^-
\end{equation}
of an initially $q$-fold charged atomic ion $A^{q+}$, the electron energy is detuned from the cooling condition such that nonzero electron-ion collision energies are realized. Since all particles move with high velocity (typically a few percent of  the speed of light) the product ions $A^{(q-1)+}$ and $A^{(q+1)+}$ are easily detected by placing single-particle detectors at appropriate locations behind the first storage-ring dipole magnet behind the electron cooler (Fig.~\ref{fig:principle}). Thus, the merged-beam technique makes up for the diluteness of the charged-particle beams by proving a large interaction volume and a detection efficiency of practically 100\%. Merged-beams recombination and ionization rate coefficients or cross sections as a function of electron-ion collision energy are readily derived from the measured  product-ion count rates by normalizing these on electron current, number of stored primary ions, and beam overlap \citep{Kilgus1992,Linkemann1995a,Schippers2001c}.

The electron-ion merged-beams technique provides access to a wide range of collision-energies extending over several orders of magnitude from sub-meV \citep{Gwinner2000} to sub-MeV \citep{Bernhardt2011a}. Due to the kinematics of the merged-beam arrangement the experimental collision-energy spread is lowest at small collision energies. This feature can be exploited for spectroscopic studies of low-energy recombination resonances, in particular, in combination with an ultracold electron beam from a liquid-nitrogen-cooled photo cathode \citep{Wolf2006c,Lestinsky2008a}.

Ion-beam storage lifetimes depend mainly on residual gas density and on ion-velocity \citep{Grieser2012}. Generally lifetimes increase with ion velocity since atomic collision cross-sections decrease strongly with increasing collision energy. When the ion velocity becomes too low electron capture from the residual gas severely shortens the storage lifetime and additionally produces a sizeable background on the recombination detector. In magnetic storage rings, this hampers electron-ion collision experiments with low-charged heavy ions. There is no such limitation in electrostatic storage rings such as the CSR which will be able to store ions independent of their mass and thus provide access to a wide range of charged particles \citep{Krantz2011} which could not be investigated before.

\section{Atomic data for astrophysics and plasma physics}\label{sec:data}

One line of atomic physics research at heavy ion-storage rings aims at providing accurate atomic data, i.e., absolute cross sections for PR and EII, for applications in astrophysics and plasma physics \citep[for the most recent published results see][]{Mahmood2013,Hahn2014a,Bernhardt2014}. Since these activities have been repeatedly reviewed before \citep{Savin2007d, Schippers2009a, Schippers2010, Schippers2012a, Hahn2014, Krantz2014} they are only briefly mentioned here.

The atomic data needs for the modeling of nonequilibrium plasmas are vast. Most of the body of compiled data comes from theory. Collision theory for complex ions requires approximations  even with modern computer technology at hand (see also  Sec.~\ref{sec:pathways}). Therefore, benchmarking by experiment is vital for the further development of the theoretical methods. Much attention has been devoted to PR and EII of iron ions since iron is  abundant in many cosmic plasmas. Consequently, prominent iron features are commonly observed, e.g., by space-borne X-ray telescopes \citep{Paerels2003a}. So far PR and EII cross sections were measured for iron ions with charge states down to 7+ \citep{Schippers2010,Schippers2012a,Hahn2014}. Lower charge states which have an increasingly complex electronic structure and for which benchmarking is therefore particulary important will become accessible at the CSR.

\begin{figure}
\centering{\includegraphics[width=0.7\textwidth]{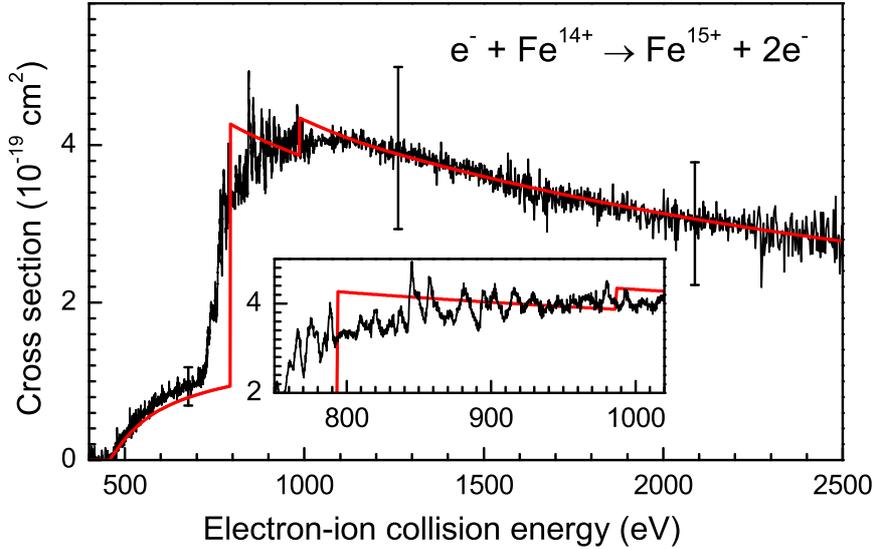}}
\caption{\label{fig:Fe} Measured cross section for electron-impact ionization of Fe$^{14+}$ ions \citep[symbols]{Bernhardt2014} in comparison with the most-recent theoretical result \citep[red full line]{Dere2007}. The capped error bars denote the systematic experimental uncertainty. The inset provides a detailed view of the strongest ionization resonances.}
\end{figure}

A recent example for such benchmarking is the cross section for EII of Fe$^{14+}$. Figure~\ref{fig:Fe} displays the measured cross section from the TSR and the theoretical cross section from the compilation of \citet{Dere2007}. There is overall agreement concerning the gross shape and the magnitude of the cross section. However, there are also significant deviations in particular in the vicinity of the calculated excitation-autoionization steps. These differences are mainly related to the neglect of resonant ionization processes in the calculation \citep[for a detailed discussion see][]{Bernhardt2014}.  For a comprehensive theoretical understanding of EII of atomic ions, resonant processes  as well as interference effects have to be included in a unified approach \citep[see, e.g.,][]{Mueller2014b} that goes beyond the widely used independent-processes approximation.

As compared to the single-pass crossed-beams technique \citep[see, e.g.,][]{Mueller2014b}  the storage-ring merged-beam technique offers additional control of the ion beam composition. In a storage-ring, metastable components that often contaminate the ion beam in single-pass experiments can relax to the ground state before data taking starts. For example, no signs of metastable $3s\,3p\;^{3\!}P$ levels were observed in the storage ring experiment with Mg-like Fe$^{14+}$ of \citet[][Fig.~\ref{fig:Fe}]{Bernhardt2014}.

\section{Exotic recombination pathways}\label{sec:pathways}

The most common resonant electron-ion recombination process is dielectronic recombination where the initially free electron excites an initially bound electron and, thereby, looses energy such it becomes bound. This initial step of the DR process is termed dielectronic capture and can be viewed as the time-inverse of the autoionization. It leads to the formation of an intermediate doubly excited state. When this intermediate state decays via photon emission the charge state of the product ion is stabilized and the DR process is completed. Accordingly, the DR cross section is proportional to the initial dielectronic capture rate  $A_a$ (autoionization rate) and to the branching ratio for radiative stabilization, i.e.,
\begin{equation}\label{eq:sigmaDR}
  \sigma \propto A_a\frac{A_r}{A_r+A_a} = \left\{\begin{array}{ll}A_r & \textrm{~~~for~}A_r\ll A_a\\A_a & \textrm{~~~for~}A_r\gg A_a\end{array}\right.
\end{equation}
where the limiting cases apply to intermediate levels with large differences between the autoionization rate $A_a$ and the radiative decay rate $A_r$.

\begin{figure}
\centering{\includegraphics[width=0.6\textwidth]{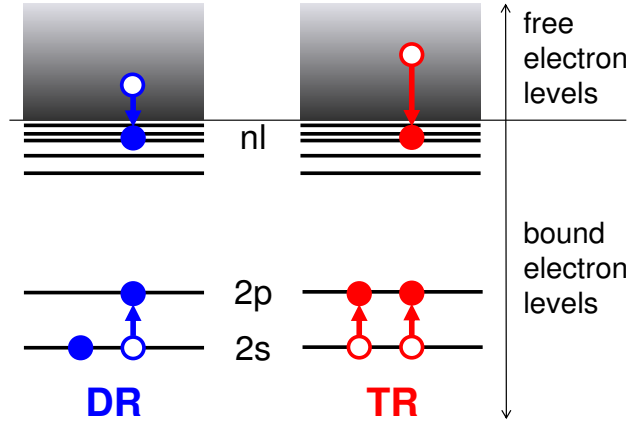}}
\caption{\label{fig:DRTR} Simplified energy-level diagrams for dielectronic recombination (DR) and trielectronic recombination (TR) of Be-like ions. The K-shell electrons are not shown.}
\end{figure}

\begin{figure}
\centering{\includegraphics[width=0.7\textwidth]{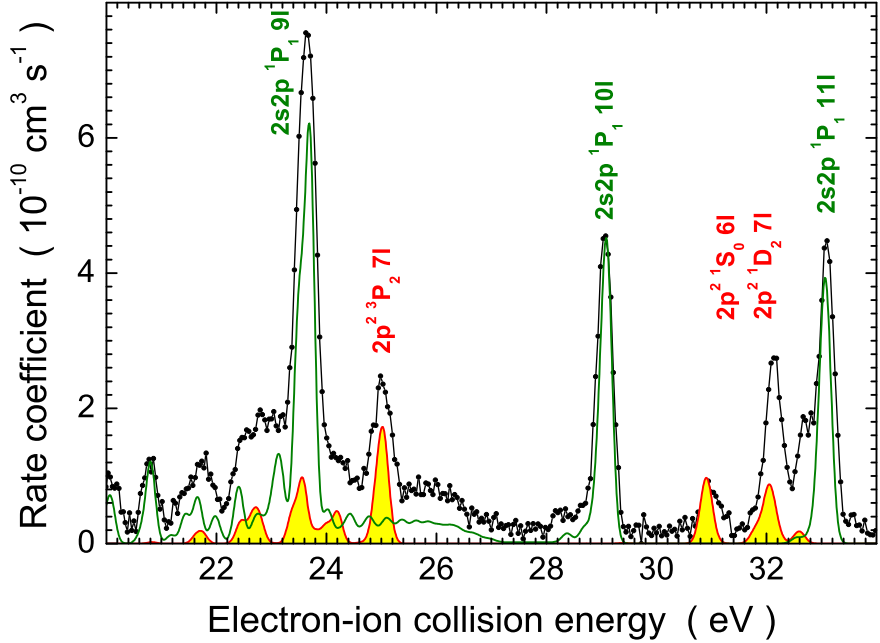}}
\caption{\label{fig:Cl13} Part of the Cl$^{13+}$ recombination spectrum measured by \citet[symbols]{Schnell2003b} at the Heidelberg heavy-ion storage ring TSR. Also shown are theoretical results for DR (green full line) and TR (red full line with yellow shading). The designations of some of the identified intermediate DR and TR resonance states are annotating the respective experimental resonance features.}
\end{figure}

In principle, one can also expect higher order processes where more than two electrons are involved. Such a process is trielectronic recombination (TR) where the incoming electron interacts simultaneously with two initially bound electrons. Figure~\ref{fig:DRTR} sketches the TR process in a Be-like ion, where the interaction with the incoming electron may lead to a $2s^2\to2p^2$ core double excitation. TR resonances have first been identified by \citet[Fig.~\ref{fig:Cl13}]{Schnell2003b} in a storage-ring recombination measurement with Be-like Cl$^{13+}$ ions. It was found that the contribution of TR to the total recombination rate coefficient is substantial (up to 40\%, depending on plasma temperature). This finding, which evidently is relevant for astrophysics has been confirmed by storage-ring measurements with other Be-like ions \citep{Schippers2004c,Fogle2005a,Savin2006a,Orban2010,Ali2013}. TR has also been observed in electron-ion recombination experiments in an electron-beam ion trap \citep[EBIT,][]{Beilmann2011a,Beilmann2013}. In these measurements weak signatures of the next-higher-order process, i.e., quadelectronic recombination (QR), were observed as well.

On the theoretical side, higher-order recombination processes may be included by allowing for configuration interactions in the calculations \citep{Schnell2003b,Colgan2003a,Beilmann2013}. This procedure yields satisfactory results for few-electron ions. For complex many-electron systems, however, the configuration expansions required to incorporate all relevant recombination pathways become prohibitively large. An example, for such a system is the open-$4f$-shell ion  Au$^{25+}$([Kr]$\,4d^{10}\,4f^8$). Experimentally, recombination of Au$^{25+}$ has been studied by \citet{Hoffknecht1998} who used a single-pass electron-ion merged-beam approach at the UNILAC linear accelerator at GSI in Darmstadt, Germany. In these experiments an exceptionally large recombination rate coefficient was observed at zero electron-ion collision energy, exceeding the expectation for nonresonant radiative recombination (RR) by more than two orders of magnitude. It was concluded that this excessive recombination rate coefficient is due to mutually overlapping strong recombination resonances.

\begin{figure}[ttt]
\centering{\includegraphics[width=0.8\textwidth]{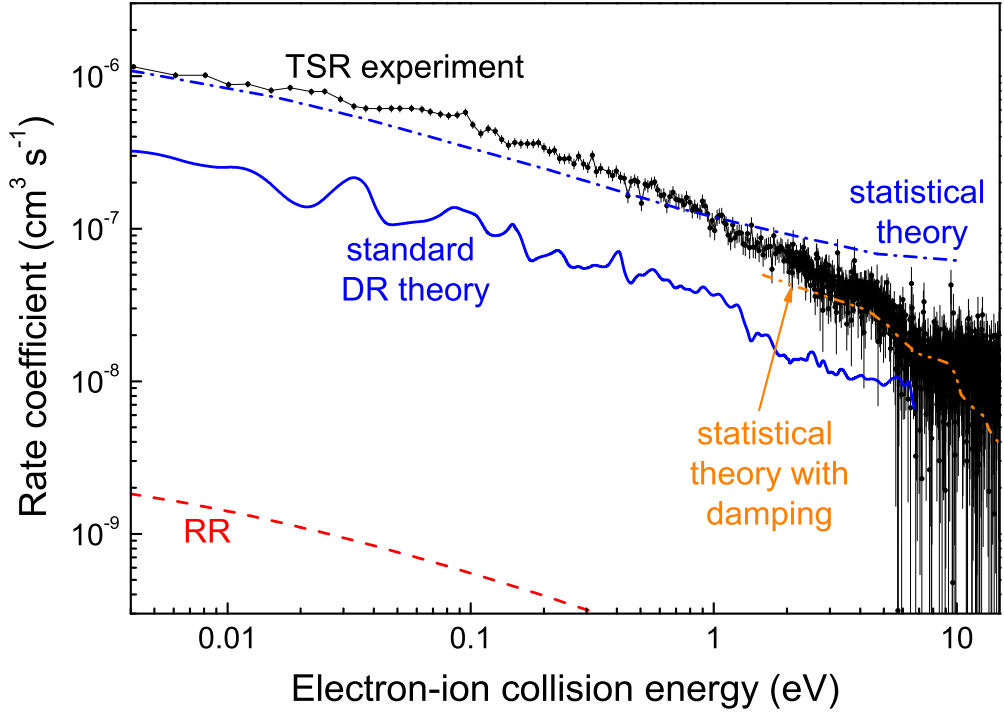}}
\caption{\label{fig:W20} Merged-beams rate coefficients for recombination of W$^{20+}$([Kr]$\,4d^{10}\,4f^8$) ions with free electrons. Symbols denote experimental results of \citet{Schippers2011} from the TSR storage ring. In the displayed energy range it is almost three orders of magnitude larger than the theoretical rate coefficient for radiative recombination \citep[RR, red dashed line, from ][]{Schippers2011}. The blue full line and the blue dash-dotted line are the results of standard DR theory and of undamped statistical theory, respectively, from \citet{Badnell2012}. The orange dash-dot-dotted line is the results of the damped statistical theory of \citet{Dzuba2013}. It should be noted that accurate rate coefficients for the recombination of tungsten ions are of immediate interest for the fusion community, since tungsten is the material of choice for plasma facing components in the ITER tokamak \citep{Pitts2013}. Accurate atomic data are required for the modeling of radiation losses from the fusion plasma due to impurity ions, and recombination rate coefficients for tungsten ions are of particular concern \citep{Puetterich2008}.
For W$^{20+}$ \citep{Schippers2011,Krantz2014} and W$^{18+}$ \citep{Spruck2014a} it was found that in the temperature ranges where these ions form in a collisionally ionized plasma the experimentally derived plasma recombination rate coefficients exceed the ones from the ADAS atomic data base \citep{ADAS} by up to an order of magnitude.}
\end{figure}

Subsequent theoretical investigations revealed a large density of multiply excited Au$^{24+}$ resonance levels. The corresponding spacing between levels of same angular momentum and parity was estimated to be about 1~meV for resonance energies below 1 eV \citep{Gribakin1999}. As already mentioned, the explicit inclusion of all these resonance levels in the configuration expansions is not possible due to practical limitations. In this situation \citet{Flambaum2002} have applied statistical theory to electron ion recombination and achieved good agreement with the experimental low-energy recombination rate coefficients for Au$^{25+}$ of \citet{Hoffknecht1998}.

In statistical theory \citep{Flambaum2002}, only a limited number of \lq\lq doorway\rq\rq\  resonance levels are considered in the initial dielectronic capture process. Additional resonance levels associated with many-electron interactions are taken into account by mixing, i.e., by performing a weighted average over many adjacent densely spaced levels of the same symmetry from a statistical Breit-Wigner distribution. The mixed-in multiply excited levels are not directly coupled to the target continuum. Therefore, the autoionization rate $A_a^{(m)}\ll A_r$ of the mixed level is smaller than the autoionization rate $A_a^{(d)}\gg A_r$ of the doorway level. Consequently, the resonance cross section (Eq.\ref{eq:sigmaDR}) changes from $\sigma \propto A_a^{(d)}{A_r}/{(A_r+A_a^{(d)})} \propto A_r$ to $\sigma \propto A_a^{(d)}{A_r}/{(A_r+A_a^{(m)})} \propto A_a^{(d)}$. Thus, the mixing of the statistically distributed levels increases the resonance cross section by a factor $A_a^{(d)}/{A_r}$ as explained in more detail by \citet{Badnell2012}.

Figure~\ref{fig:W20} summarizes the state-of-the-art for electron-ion recombination of W$^{20+}$([Kr]$\,4d^{10}\,4f^8$). As in the case of isoelectronic Au$^{25+}$,  the measured merged-beam rate coefficient \citep{Schippers2011} at low electron-ion collision energies is more than two orders of magnitude larger than the theoretical rate coefficient for RR. Up to energies below $\sim 10$~eV the result of the standard DR calculation of \citet[][full curve]{Badnell2012} is a factor of three lower than the experimental curve. When statistical mixing is included the theoretical result agrees with the experimental low-energy rate coefficient \citep[][dash-dotted curve]{Badnell2012}. A similar agreement between experiment and statistical theory for W$^{20+}$ was also found by \citet{Dzuba2012}. However, both theoretical results overestimate the experimental rate coefficient at energies above $\sim 1$~eV. This is due to the neglect of autoionization of resonance states to excited final levels. These deexcitation channels open up at higher electron-ion collision energies and lead to a reduction of the fluorescence yield. When including this damping effect into statistical theory good agreement with the experimental result is also achieved at higher electron-ion collision energies \citep[][dash-dot-dotted curve in Fig.~\ref{fig:W20}]{Dzuba2013}. Even better agreement between experiment and statistical theory with damping has been recently established for recombination of  W$^{18+}$([Kr]$\,4d^{10}\,4f^{10}$) \citep{Spruck2014a}. Clearly, the fruitful interplay between experiment and theory has advanced our understanding of electron-ion recombination of complex ions.

Complex atomic and molecular systems also exhibit collective phenomena. For example the cross section for photoionization of xenon atoms \citep{West1978} is dominated by a \lq\lq giant\rq\rq\ resonance that is associated with the collective excitation of all $4d$ electrons. Since photoionization is the time inverse of photorecombination, recombination may also proceed via collective excitations. This recombination pathway can be expected to be significant in low-charged heavy ions similar to the process of polarization recombination \citep{Bureyeva1998,Korol2006a}. As already mentioned above, recombination studies with such ions will become feasible in the Heidelberg CSR storage ring. In addition, even more heavy species will become accessible such as (endohedral) fullerene ions which also exhibit strong collective photoionization resonances \cite[][and references therein]{Scully2005a,Phaneuf2013a}.

\section{Electron-ion collision spectroscopy of highly charged ions}\label{sec:spec}

In general, recombination resonance features can be exploited for atomic spectroscopy of multiply excited states. It should be noted that this experimental technique, which may be referred to as \lq\lq electron-ion collision spectroscopy\rq\rq\ does not involve the excitation by or the detection of any photon \citep[see also Sec.~\ref{sec:exp}]{Brandau2010}. Recent related experiments \citep[see also][]{Schippers2009} comprise the determination of hyperfine-structure splittings \citep{Lindroth2001,Lestinsky2008a} and nuclear charge radii \citep{Brandau2008a} in heavy few electron systems, the quantitative investigation of the influence of the Breit interaction on recombination resonance strength in H-like uranium \citep{Bernhardt2011a}, and the spectroscopy of ions with in-flight produced unstable nuclei \citep{Brandau2013}. Such spectroscopy studies will greatly benefit from the installation of \textsc{Cryring} at the international FAIR facility in Darmstadt, Germany. \textsc{Cryring} will be coupled to the existing ESR storage-ring \citep{Stoehlker2014} which will serve as an injector of precooled and decelerated  ions. The combination of excellent vacuum conditions with a cold electron beam will facilitate high-resolution spectroscopic studies of highly charged heavy ions and, thus, the investigation of relativistic, QED, and nuclear effects in such systems with unprecedented precision. In addition, the experimental investigation of resonant processes involving nuclear excitations \citep{Palffy2010} may become feasible.

\begin{figure}[t]
\centering{\includegraphics[width=0.8\columnwidth]{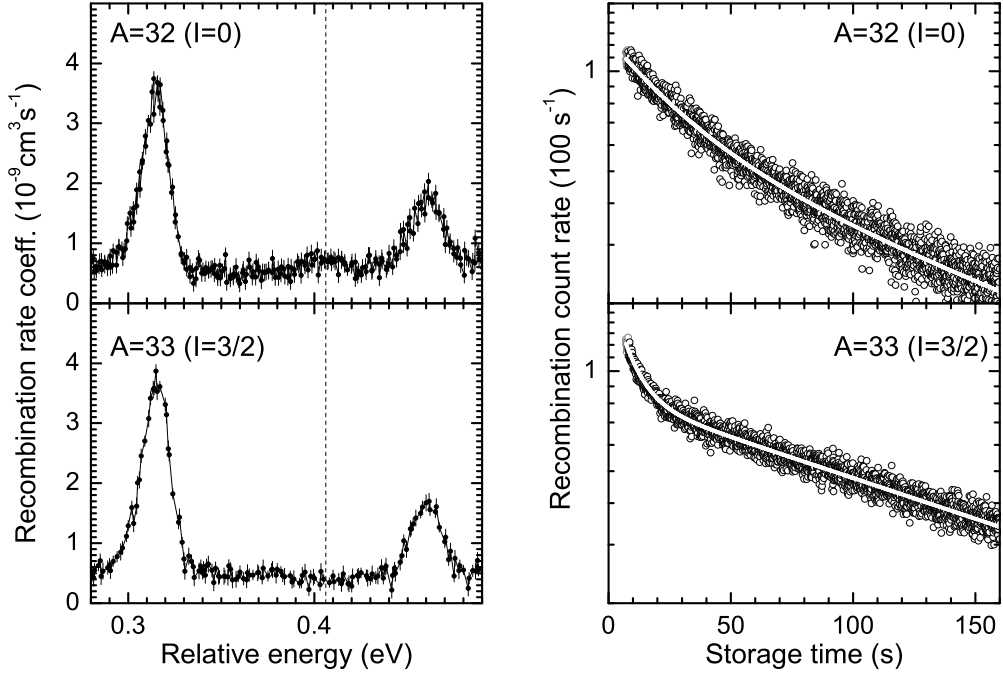}}
\caption{\label{fig:S12lifetime}Left: Measured recombination spectra \citep{Schippers2012}
for $^{32}$S$^{12+}$ and $^{33}$S$^{12+}$ ions, with nuclear sopins $I=0$ and $I=3/2$, respectively. A dielectronic recombination resonance at $\sim$0.4~eV (vertical dashed line) that is excited from
the metastable $1s^2\,2s\,2p\;^3P_0$ level  is not visible in the $^{33}$S$^{12+}$ spectrum, since
the $^3P_0$ state is quenched by the hyperfine interaction after sufficiently long storage times.
Only resonances that are excited from the $1s^2\,2s^2\;^1S_0$ ground level occur in both spectra. Right:
Measured (symbols) and fitted (lines) decay curves \citep{Schippers2012} for $^{32}$S$^{12+}$
and $^{33}$S$^{12+}$ ions stored in the TSR heavy-ion storage ring.
The monitored signal is the S$^{11+}$ production rate at an electron-ion collision energy of 0.406~eV
(marked by the vertical dashed line in the left panel). In contrast to the $A=32$ (upper) curve the $A=33$ (lower)
curve exhibits a fast decaying component which is the signature of the hyperfine induced $2s\,2p\;^3P_0 \to 2s^2\;^1S_0$ transition. For this transition a transition rate of 0.096(4)~s$^{-1}$ was derived from the measured decay curves \citep{Schippers2012}. The figure has been taken from \citep{Schippers2013}.}
\end{figure}

Electron-ion collision spectroscopy can also be used for measuring atomic transition rates. An example is the measurement of hyperfine induced transition rates in Be-like ions \citep{Schippers2007a,Schippers2012}. Here electron-ion recombination spectroscopy is the only method that provided experimental access to these exotic atomic transitions in multiply and highly charged ions \citep{Schippers2013}. In these experiments the decay of the number of stored ions is monitored by employing electron-ion recombination (Fig.~\ref{fig:S12lifetime}). In order to selectively enhance the signal from the metastable $2s\,2p\;^3P_0$ level, the electron-ion collision energy is tuned to a dielectronic recombination resonance associated with this level. An essential feature of this experimental method is the comparison of measured results from different isotopes with zero and nonzero nuclear spin. The method is readily applicable to a wide range of ions. In the future, beams of radioactive nuclei from HIE-ISOLDE may become available at the TSR storage ring \citep{Grieser2012}. Measurements of HFI induced transition rates may then be used for determining nuclear magnetic moments, some of which are still known only imprecisely. In addition, the method may also be used to study other weak atomic decay channels such as two-photon transitions \citep{Bernhardt2012b}.

\section*{Acknowledgments}

The author is indebted to many colleagues who have contributed to the featured research and whose names appear in the list of references below. The author is particularly thankful to Claude Krantz, Alfred M\"{u}ller, Daniel Wolf Savin and Andreas Wolf for long-standing fruitful scientific collaboration at the TSR storage ring. The featured research has in part be funded by the Deutsche Forschungsgemeinschaft (DFG, contract numbers Schi378/8-1 and Schi378/9-1).


\begin{thebibliography}{62}
\expandafter\ifx\csname natexlab\endcsname\relax\def\natexlab#1{#1}\fi
\expandafter\ifx\csname url\endcsname\relax
  \def\url#1{\texttt{#1}}\fi
\expandafter\ifx\csname urlprefix\endcsname\relax\def\urlprefix{}\fi

\bibitem[{ADAS(2014)}]{ADAS}
ADAS, 2014. {Atomic Data and Analysis Structure}.
\urlprefix\url{http://www.adas.ac.uk}

\bibitem[{{Ali} et~al.(2013){Ali}, {Orban}, {Mahmood}, {Loch}, and
  {Schuch}}]{Ali2013}
{Ali}, S., {Orban}, I., {Mahmood}, S., {Loch}, S.~D., {Schuch}, R., Sep. 2013.
  {Experimental rate coefficients of F$^{5+}$ recombining into F$^{4+}$}.
  Astron. Astrophys. 557, A2.

\bibitem[{Badnell et~al.(2012)Badnell, Ballance, Griffin, and
  O'Mullane}]{Badnell2012}
Badnell, N.~R., Ballance, C.~P., Griffin, D.~C., O'Mullane, M., May 2012.
  Dielectronic recombination of {W}$^{20+}$ ($4{d}^{10}4{f}^{8}$): Addressing
  the half-open $f$ shell. Phys. Rev. A 85, 052716.


\bibitem[{Beilmann et~al.(2013)Beilmann, Harman, Mokler, Bernitt, Keitel,
  Ullrich, and {Crespo L\'opez-Urrutia}}]{Beilmann2013}
Beilmann, C., Harman, Z., Mokler, P.~H., Bernitt, S., Keitel, C.~H., Ullrich,
  J., {Crespo L\'opez-Urrutia}, J.~R., 2013. Major role of multielectronic
  {K}-{L} intershell resonant recombination processes in {L}i- to {O}-like ions
  of {A}r, {F}e, and {K}r. Phys. Rev. A 88, 062706.

\bibitem[{Beilmann et~al.(2011)Beilmann, Mokler, Bernitt, Keitel, Ullrich,
  {Crespo L\'opez-Urrutia}, and Harman}]{Beilmann2011a}
Beilmann, C., Mokler, P.~H., Bernitt, S., Keitel, C.~H., Ullrich, J., {Crespo
  L\'opez-Urrutia}, J.~R., Harman, Z., 2011. Prominent higher-order
  contributions to electronic recombination. Phys. Rev. Lett. 107, 143201.

\bibitem[{Bernhardt et~al.(2014)Bernhardt, Becker, Grieser, Hahn, Krantz,
  Lestinsky, Novotn\'y, Repnow, Savin, Spruck, Wolf, M\"uller, and
  Schippers}]{Bernhardt2014}
Bernhardt, D., Becker, A., Grieser, M., Hahn, M., Krantz, C., Lestinsky, M.,
  Novotn\'y, O., Repnow, R., Savin, D.~W., Spruck, K., Wolf, A., M\"uller, A.,
  Schippers, S., 2014. Absolute rate coefficients for photorecombination and
  electron-impact ionization of magnesiumlike iron ions from measurements at a
  heavy-ion storage ring. Phys. Rev. A 90, 012702.

\bibitem[{Bernhardt et~al.(2011)Bernhardt, Brandau, Harman, Kozhuharov,
  M\"uller, Scheid, Schippers, Schmidt, Yu, Artemyev, Tupitsyn, B\"ohm, Bosch,
  Currell, Franzke, Gumberidze, Jacobi, Mokler, Nolden, Spillman, Stachura,
  Steck, and St\"ohlker}]{Bernhardt2011a}
Bernhardt, D., Brandau, C., Harman, Z., Kozhuharov, C., M\"uller, A., Scheid,
  W., Schippers, S., Schmidt, E.~W., Yu, D., Artemyev, A.~N., Tupitsyn, I.~I.,
  B\"ohm, S., Bosch, F., Currell, F.~J., Franzke, B., Gumberidze, A., Jacobi,
  J., Mokler, P.~H., Nolden, F., Spillman, U., Stachura, Z., Steck, M.,
  St\"ohlker, T., 2011. Breit interaction in dielectronic recombination of
  hydrogenlike uranium. Phys. Rev. A 83, 020701.

\bibitem[{Bernhardt et~al.(2012)Bernhardt, Brandau, Kozhuharov, M\"{u}ller,
  Schippers, B\"{o}hm, Bosch, Jacobi, Kieslich, Knopp, Mokler, Nolden, Shi,
  Stachura, Steck, and St\"{o}hlker}]{Bernhardt2012b}
Bernhardt, D., Brandau, C., Kozhuharov, C., M\"{u}ller, A., Schippers, S.,
  B\"{o}hm, S., Bosch, F., Jacobi, J., Kieslich, S., Knopp, H., Mokler, P.~H.,
  Nolden, F., Shi, W., Stachura, Z., Steck, M., St\"{o}hlker, T., 2012. Towards
  a measurement of the $2s\,2p\;^3p_0 \to 2s^2\;^1s_0$ {E1M1} two photon
  transition rate in {B}e-like xenon ions. J. Phys.: Conf. Ser. 388, 012007.


\bibitem[{Brandau et~al.(2008)Brandau, Kozhuharov, Harman, M\"{u}ller,
  Schippers, Kozhedub, Bernhardt, B\"{o}hm, Jacobi, Schmidt, Mokler, Bosch,
  Kluge, St\"{o}hlker, Beckert, Beller, Nolden, Steck, Gumberidze, Reuschl,
  Spillmann, Currell, Tupitsyn, Shabaev, Jentschura, Keitel, Wolf, and
  Stachura}]{Brandau2008a}
Brandau, C., Kozhuharov, C., Harman, Z., M\"{u}ller, A., Schippers, S.,
  Kozhedub, Y.~S., Bernhardt, D., B\"{o}hm, S., Jacobi, J., Schmidt, E.~W.,
  Mokler, P.~H., Bosch, F., Kluge, H.-J., St\"{o}hlker, T., Beckert, K.,
  Beller, P., Nolden, F., Steck, M., Gumberidze, A., Reuschl, R., Spillmann,
  U., Currell, F., Tupitsyn, I.~I., Shabaev, V.~M., Jentschura, U.~D., Keitel,
  C.~H., Wolf, A., Stachura, Z., 2008. Isotope shift in the dielectronic
  recombination of three-electron $^{A}${N}d$^{57+}$. Phys. Rev. Lett. 100,
  073201.

\bibitem[{Brandau et~al.(2013)Brandau, Kozhuharov, M\"{u}ller, Bernhardt,
  Banas, Bosch, Currell, Dimopoulou, Gumberidze, Hagmann, Hillenbrand, Heil,
  Lestinsky, Litvinov, M\"{a}rtin, Nolden, Reuschl, Sanjari, Schippers,
  Schneider, Shubina, Simon, Spillmann, Stachura, Steck, St\"{o}hlker, Weber,
  Wiedeking, Winckler, and Winters}]{Brandau2013}
Brandau, C., Kozhuharov, C., M\"{u}ller, A., Bernhardt, D., Banas, D., Bosch,
  F., Currell, F.~J., Dimopoulou, C., Gumberidze, A., Hagmann, S., Hillenbrand,
  P.-M., Heil, M., Lestinsky, M., Litvinov, Y.~A., M\"{a}rtin, R., Nolden, F.,
  Reuschl, R., Sanjari, S., Schippers, S., Schneider, D., Shubina, D., Simon,
  H., Spillmann, U., Stachura, Z., Steck, M., St\"{o}hlker, T., Weber, G.,
  Wiedeking, M., Winckler, N., Winters, D. F.~A., 2013. Probing nuclear
  properties by resonant atomic collisions between electrons and ions. Phys.
  Scr. T156, 014050.

\bibitem[{Brandau et~al.(2010)Brandau, Kozhuharov, M\"{u}ller, Bernhardt,
  Bosch, Boutin, Currell, Dimopoulou, Franzke, Fritzsche, Gumberidze, Harman,
  Jentschura, Keitel, Kozhedub, Kr\"{u}cken, Litvinov, Nolden, {O' Rourke},
  Reuschl, Schippers, Shabaev, Spillmann, Stachura, Steck, St\"{o}hlker,
  Tupitsyn, Winters, and Wolf}]{Brandau2010}
Brandau, C., Kozhuharov, C., M\"{u}ller, A., Bernhardt, D., Bosch, F., Boutin,
  D., Currell, F., Dimopoulou, C., Franzke, B., Fritzsche, S., Gumberidze, A.,
  Harman, Z., Jentschura, U., Keitel, C., Kozhedub, Y., Kr\"{u}cken, R.,
  Litvinov, Y., Nolden, F., {O' Rourke}, B., Reuschl, R., Schippers, S.,
  Shabaev, V., Spillmann, U., Stachura, Z., Steck, M., St\"{o}hlker, T.,
  Tupitsyn, I., Winters, D., Wolf, A., 2010. Resonant recombination at ion
  storage rings: a conceptual alternative for isotope shift and hyperfine
  studies. Hyperfine Interact. 196, 115.

\bibitem[{Bureyeva and Lisitsa(1998)}]{Bureyeva1998}
Bureyeva, L., Lisitsa, V., 1998. Polarization recombination as a new channel
  for recombination of free electrons on complex ions. J. Phys. B 31,
  1477.

\bibitem[{Colgan et~al.(2003)Colgan, Pindzola, Whiteford, and
  Badnell}]{Colgan2003a}
Colgan, J., Pindzola, M.~S., Whiteford, A.~D., Badnell, N.~R., 2003.
  Dielectronic recombination data for dynamic finite-density plasmas {III}.
  {T}he beryllium isoelectronic sequence. Astron. Astrophys. 412, 597.

\bibitem[{Dere(2007)}]{Dere2007}
Dere, K.~P., 2007. Ionization rate coefficients for the elements hydrogen
  through zinc. Astron. Astrophys. 466, 771.

\bibitem[{Dzuba et~al.(2012)Dzuba, Flambaum, Gribakin, and
  Harabati}]{Dzuba2012}
Dzuba, V.~A., Flambaum, V.~V., Gribakin, G.~F., Harabati, C., 2012.
  Chaos-induced enhancement of resonant multielectron recombination in highly
  charged ions: Statistical theory. Phys. Rev. A 86, 022714.

\bibitem[{Dzuba et~al.(2013)Dzuba, Flambaum, Gribakin, Harabati, and
  Kozlov}]{Dzuba2013}
Dzuba, V.~A., Flambaum, V.~V., Gribakin, G.~F., Harabati, C., Kozlov, M.~G.,
  2013. Electron recombination, photoionization, and scattering via
  many-electron compound resonances. Phys. Rev. A 88, 062713.

\bibitem[{Flambaum et~al.(2002)Flambaum, Gribakina, Gribakin, and
  Harabati}]{Flambaum2002}
Flambaum, V.~V., Gribakina, A.~A., Gribakin, G.~F., Harabati, C., 2002.
  Electron recombination with multicharged ions via chaotic many-electron
  states. Phys. Rev. A 66, 012713.

\bibitem[{Fogle et~al.(2005)Fogle, Badnell, Glans, Loch, Madzunkov, Abdel-Naby,
  Pindzola, and Schuch}]{Fogle2005a}
Fogle, M., Badnell, N.~R., Glans, P., Loch, S.~D., Madzunkov, S., Abdel-Naby,
  S.~A., Pindzola, M.~S., Schuch, R., 2005. Electron-ion recombination of
  {B}e-like {C}, {N}, and {O}. Astron. Astrophys. 442, 757.

\bibitem[{Gribakin et~al.(1999)Gribakin, Gribakina, and
  Flambaum}]{Gribakin1999}
Gribakin, G.~F., Gribakina, A.~A., Flambaum, V.~V., 1999. Quantum chaos in
  multicharged ions and statistical approach to the calculation of electron-ion
  resonant radiative recombination. Aust. J. Phys. 52, 443.

\bibitem[{Grieser et~al.(2012)Grieser, Litvinov, Raabe, Blaum, Blumenfeld,
  Butler, Wenander, Woods, Aliotta, Andreyev, Artemyev, Atanasov, Aumann,
  Balabanski, Barzakh, Batist, Bernardes, Bernhardt, Billowes, Bishop, Borge,
  Borzov, Bosch, Boston, Brandau, Catford, Catherall, Cederk\"{a}ll, Cullen,
  Davinson, Dillmann, Dimopoulou, Dracoulis, D\"{u}llmann, Egelhof, Estrade,
  Fischer, Flanagan, Fraile, Fraser, Freeman, Geissel, Gerl, Greenlees,
  Grisenti, Habs, von Hahn, Hagmann, Hausmann, He, Heil, Huyse, Jenkins,
  Jokinen, Jonson, Joss, Kadi, Kalantar-Nayestanaki, Kay, Kiselev, Kluge,
  Kowalska, Kozhuharov, Kreim, Kr\"{o}ll, Kurcewicz, Labiche, Lemmon,
  Lestinsky, Lotay, Ma, Marta, Meng, M\"{u}cher, Mukha, M\"{u}ller, Murphy,
  Neyens, Nilsson, Nociforo, N\"{o}rtersh\"{a}user, Page, Pasini, Petridis,
  Pietralla, Pf\"{u}tzner, Podoly\'{a}k, Regan, Reed, Reifarth, Reiter, Repnow,
  Riisager, Rubio, Sanjari, Savin, Scheidenberger, Schippers, Schneider,
  Schuch, Schwalm, Schweikhard, Shubina, Siesling, Simon, Simpson, Smith,
  Sonnabend, Steck, Stora, St\"{o}hlker, Sun, Surzhykov, Suzaki, Tarasov,
  Trotsenko, Tu, Van~Duppen, Volpe, Voulot, Walker, Wildner, Winckler, Winters,
  Wolf, Xu, Yakushev, Yamaguchi, Yuan, Zhang, and Zuber}]{Grieser2012}
Grieser, M., Litvinov, Y., Raabe, R., Blaum, K., Blumenfeld, Y., Butler, P.~A.,
  Wenander, F., Woods, P.~J., Aliotta, M., Andreyev, A., Artemyev, A.,
  Atanasov, D., Aumann, T., Balabanski, D., Barzakh, A., Batist, L., Bernardes,
  A.-P., Bernhardt, D., Billowes, J., Bishop, S., Borge, M., Borzov, I., Bosch,
  F., Boston, A.~J., Brandau, C., Catford, W., Catherall, R., Cederk\"{a}ll,
  J., Cullen, D., Davinson, T., Dillmann, I., Dimopoulou, C., Dracoulis, G.,
  D\"{u}llmann, C.~E., Egelhof, P., Estrade, A., Fischer, D., Flanagan, K.,
  Fraile, L., Fraser, M.~A., Freeman, S.~J., Geissel, H., Gerl, J., Greenlees,
  P., Grisenti, R.~E., Habs, D., von Hahn, R., Hagmann, S., Hausmann, M., He,
  J.~J., Heil, M., Huyse, M., Jenkins, D., Jokinen, A., Jonson, B., Joss,
  D.~T., Kadi, Y., Kalantar-Nayestanaki, N., Kay, B.~P., Kiselev, O., Kluge,
  H.-J., Kowalska, M., Kozhuharov, C., Kreim, S., Kr\"{o}ll, T., Kurcewicz, J.,
  Labiche, M., Lemmon, R.~C., Lestinsky, M., Lotay, G., Ma, X.~W., Marta, M.,
  Meng, J., M\"{u}cher, D., Mukha, I., M\"{u}ller, A., Murphy, A. S.~J.,
  Neyens, G., Nilsson, T., Nociforo, C., N\"{o}rtersh\"{a}user, W., Page,
  R.~D., Pasini, M., Petridis, N., Pietralla, N., Pf\"{u}tzner, M.,
  Podoly\'{a}k, Z., Regan, P., Reed, M.~W., Reifarth, R., Reiter, P., Repnow,
  R., Riisager, K., Rubio, B., Sanjari, M.~S., Savin, D.~W., Scheidenberger,
  C., Schippers, S., Schneider, D., Schuch, R., Schwalm, D., Schweikhard, L.,
  Shubina, D., Siesling, E., Simon, H., Simpson, J., Smith, J., Sonnabend, K.,
  Steck, M., Stora, T., St\"{o}hlker, T., Sun, B., Surzhykov, A., Suzaki, F.,
  Tarasov, O., Trotsenko, S., Tu, X.~L., Van~Duppen, P., Volpe, C., Voulot, D.,
  Walker, P.~M., Wildner, E., Winckler, N., Winters, D. F.~A., Wolf, A., Xu,
  H.~S., Yakushev, A., Yamaguchi, T., Yuan, Y.~J., Zhang, Y., Zuber, K., 2012.
  Storage ring at {HIE-ISOLDE}. Eur. Phys. J. ST 207, 1.

\bibitem[{Gwinner et~al.(2000)Gwinner, Hoffknecht, Bartsch, Beutelspacher,
  Ekl\"{o}w, Glans, Grieser, Krohn, Lindroth, M\"{u}ller, Saghiri, Schippers,
  Schramm, Schwalm, Tokman, Wissler, and Wolf}]{Gwinner2000}
Gwinner, G., Hoffknecht, A., Bartsch, T., Beutelspacher, M., Ekl\"{o}w, N.,
  Glans, P., Grieser, M., Krohn, S., Lindroth, E., M\"{u}ller, A., Saghiri,
  A.~A., Schippers, S., Schramm, U., Schwalm, D., Tokman, M., Wissler, G.,
  Wolf, A., 2000. Influence of magnetic fields on electron-ion recombination at
  very low energies. Phys. Rev. Lett. 84, 4822.

\bibitem[{{Hahn}(2014)}]{Hahn2014}
{Hahn}, M., 2014. Electron impact ionization of stored highly charged ions. J.
  Phys.: Conf. Ser. 488, 012050.

\bibitem[{Hahn et~al.(2014)Hahn, Badnell, Grieser, Krantz, Lestinsky,
  M\"{u}ller, Novotn\'{y}, Repnow, Schippers, Wolf, and Savin}]{Hahn2014a}
Hahn, M., Badnell, N.~R., Grieser, M., Krantz, C., Lestinsky, M., M\"{u}ller,
  A., Novotn\'{y}, O., Repnow, R., Schippers, S., Wolf, A., Savin, D.~W., 2014.
  Electron-ion recombination of {F}e$^{12+}$ forming {F}e$^{11+}$: laboratory
  measurements and theoretical calculations. Astrophys. J. 788, 46.

\bibitem[{Hoffknecht et~al.(1998)Hoffknecht, Uwira, Frank, Schennach, Spies,
  Wagner, Schippers, M\"{u}ller, Becker, Kleinod, Angert, and
  Mokler}]{Hoffknecht1998}
Hoffknecht, A., Uwira, O., Frank, A., Schennach, S., Spies, W., Wagner, M.,
  Schippers, S., M\"{u}ller, A., Becker, R., Kleinod, M., Angert, N., Mokler,
  P.~H., 1998. Recombination of {A}u$^{25+}$ with free electrons at very low
  energies. J. Phys. B 31, 2415.

\bibitem[{Kilgus et~al.(1992)Kilgus, Habs, Schwalm, Wolf, Badnell, and
  M\"{u}ller}]{Kilgus1992}
Kilgus, G., Habs, D., Schwalm, D., Wolf, A., Badnell, N.~R., M\"{u}ller, A.,
  1992. High-resolution measurement of dielectronic recombination of
  lithiumlike {C}u$^{26+}$. Phys. Rev. A 46, 5730.

\bibitem[{Korol et~al.(2006)Korol, Gribakin, and Currell}]{Korol2006a}
Korol, A., Gribakin, G.~F., Currell, F.~J., 2006. Effect of target polarization
  in electron-ion recombination. Phys. Rev. Lett. 97, 223201.

\bibitem[{Krantz et~al.(2011)Krantz, Berg, Blaum, Fellenberger, Froese,
  Grieser, {von Hahn}, Lange, Laux, Menk, Repnow, Shornikov, and
  Wolf}]{Krantz2011}
Krantz, C., Berg, F., Blaum, K., Fellenberger, F., Froese, M., Grieser, M.,
  {von Hahn}, R., Lange, M., Laux, F., Menk, S., Repnow, R., Shornikov, A.,
  Wolf, A., 2011. The {C}ryogenic {S}torage {R}ing and its application to
  molecular ion recombination physics. J. Phys.: Conf. Ser. 300, 012010.

\bibitem[{Krantz et~al.(2014)Krantz, Spruck, Badnell, Becker, Bernhardt,
  Grieser, Hahn, Novotn\'{y}, Repnow, Savin, Wolf, M\"{u}ller, and
  Schippers}]{Krantz2014}
Krantz, C., Spruck, K., Badnell, N.~R., Becker, A., Bernhardt, D., Grieser, M.,
  Hahn, M., Novotn\'{y}, O., Repnow, R., Savin, D.~W., Wolf, A., M\"{u}ller,
  A., Schippers, S., 2014. Absolute rate coefficients for the recombination of
  open-f-shell tungsten ions. J. Phys.: Conf. Ser. 488, 012051.

\bibitem[{Lestinsky et~al.(2008)Lestinsky, Lindroth, Orlov, Schmidt, Schippers,
  B\"{o}hm, Brandau, Sprenger, Terekhov, M\"{u}ller, and Wolf}]{Lestinsky2008a}
Lestinsky, M., Lindroth, E., Orlov, D.~A., Schmidt, E.~W., Schippers, S.,
  B\"{o}hm, S., Brandau, C., Sprenger, F., Terekhov, A.~S., M\"{u}ller, A.,
  Wolf, A., 2008. Screened radiative corrections from hyperfine-split
  dielectronic resonances in lithiumlike scandium. Phys. Rev. Lett. 100,
  033001.

\bibitem[{Lindroth et~al.(2001)Lindroth, Danared, Glans, Pesic, Tokman, Vikor,
  and Schuch}]{Lindroth2001}
Lindroth, E., Danared, H., Glans, P., Pesic, Z., Tokman, M., Vikor, G., Schuch,
  R., 2001. {QED} effects in {C}u-like {P}b recombination resonances near
  threshold. Phys. Rev. Lett. 86, 5027.

\bibitem[{Linkemann et~al.(1995)Linkemann, M\"{u}ller, Kenntner, Habs, Schwalm,
  Wolf, Badnell, and Pindzola}]{Linkemann1995a}
Linkemann, J., M\"{u}ller, A., Kenntner, J., Habs, D., Schwalm, D., Wolf, A.,
  Badnell, N.~R., Pindzola, M.~S., 1995. Electron-impact ionization of
  {F}e$^{15+}$ ions: {A}n ion storage ring cross section measurement. Phys.
  Rev. Lett. 74, 4173.

\bibitem[{Mahmood et~al.(2013)Mahmood, Orban, Ali, Glans, Bleda, Altun, and
  Schuch}]{Mahmood2013}
Mahmood, S., Orban, I., Ali, S., Glans, P., Bleda, E.~A., Altun, Z., Schuch,
  R., 2013. Recombination rate coefficients of boron-like {N}e. Astrophys. J.
  771, 78.

\bibitem[{M\"{u}ller(2008)}]{Mueller2008a}
M\"{u}ller, A., 2008. Electron-ion collisions: {F}undamental processes in the
  focus of applied research. Adv. At. Mol. Opt. Phys. 55, 293.

\bibitem[{M\"uller et~al.(2014)M\"uller, Borovik, Huber, Schippers, Fursa, and
  Bray}]{Mueller2014b}
M\"uller, A., Borovik, A., Huber, K., Schippers, S., Fursa, D.~V., Bray, I.,
  2014. Double-{$K$}-vacancy states in electron-impact single ionization of
  metastable two-electron {N}$^{5+}$($1s\,2s\;^3s_1$) ions. Phys. Rev. A 90,
  010701.

\bibitem[{M\"{u}ller and Wolf(1997)}]{Mueller1997c}
M\"{u}ller, A., Wolf, A., 1997. Heavy ion storage rings. In: Austin, J.~C.,
  Shafroth, S.~M. (Eds.), Accelerator-based atomic physics techniques and
  applications. AIP Press, Woodbury, p. 147.

\bibitem[{Orban et~al.(2010)Orban, Loch, B\"{o}hm, and Schuch}]{Orban2010}
Orban, I., Loch, S.~D., B\"{o}hm, S., Schuch, R., 2010. Recombination rate
  coefficients of {B}e-like {S}i. Astrophys. J. 721, 1603.

\bibitem[{Paerels and Kahn(2003)}]{Paerels2003a}
Paerels, F. B.~S., Kahn, S.~M., 2003. High-resolution {X}-ray spectroscopy with
  {C}handra and {XMM-N}ewton. Annu. Rev. Astron. Astrophys. 41, 291.

\bibitem[{Palffy(2010)}]{Palffy2010}
Palffy, A., 2010. Nuclear effects in atomic transitions. Contemp. Phys. 51,
  471--496.

\bibitem[{Phaneuf et~al.(1999)Phaneuf, Havener, Dunn, and
  M\"{u}ller}]{Phaneuf1999}
Phaneuf, R.~A., Havener, C.~C., Dunn, G.~H., M\"{u}ller, A., 1999. Merged-beams
  experiments in atomic and molecular physics. Rep. Prog. Phys. 62, 1143.

\bibitem[{Phaneuf et~al.(2013)Phaneuf, Kilcoyne, Aryal, Baral,
  Esteves-Macaluso, Thomas, Hellhund, Lomsadze, Gorczyca, Ballance, Manson,
  Hasoglu, Schippers, and M\"uller}]{Phaneuf2013a}
Phaneuf, R.~A., Kilcoyne, A. L.~D., Aryal, N.~B., Baral, K.~K.,
  Esteves-Macaluso, D.~A., Thomas, C.~M., Hellhund, J., Lomsadze, R., Gorczyca,
  T.~W., Ballance, C.~P., Manson, S.~T., Hasoglu, M.~F., Schippers, S.,
  M\"uller, A., 2013. Probing confinement resonances by photoionizing {X}e
  inside a {C}$_{60}^{+}$ molecular cage. Phys. Rev. A 88, 053402.

\bibitem[{Pitts et~al.(2013)Pitts, Carpentier, Escourbiac, Hirai, Komarov,
  Lisgo, Kukushkin, Loarte, Merola, Naik, Mitteau, Sugihara, Bazylev, and
  Stangeby}]{Pitts2013}
Pitts, R., Carpentier, S., Escourbiac, F., Hirai, T., Komarov, V., Lisgo, S.,
  Kukushkin, A., Loarte, A., Merola, M., Naik, A.~S., Mitteau, R., Sugihara,
  M., Bazylev, B., Stangeby, P., 2013. A full tungsten divertor for {ITER}:
  Physics issues and design status. J. Nucl. Mat. 438, S48.

\bibitem[{Poth(1990)}]{Poth1990}
Poth, H., 1990. Electron cooling: {T}heory, experiment, application. Phys.
  Rep. 196, 135.

\bibitem[{P\"{u}tterich et~al.(2008)P\"{u}tterich, Neu, Dux, Whiteford,
  O'Mullane, and the ASDEX Upgrade~Team}]{Puetterich2008}
P\"{u}tterich, T., Neu, R., Dux, R., Whiteford, A.~D., O'Mullane, M.~G., the
  ASDEX Upgrade~Team, 2008. Modelling of measured tungsten spectra from {ASDEX}
  {U}pgrade and predictions for {ITER}. Plasma Phys. Control. Fusion 50,
  085016.

\bibitem[{Savin(2007)}]{Savin2007d}
Savin, D.~W., 2007. Can heavy ion storage rings contribute to our understanding
  of the charge state distributions in cosmic atomic plasmas? J. Phys.: Conf.
  Ser. 88, 012071.

\bibitem[{Savin et~al.(2006)Savin, Gwinner, Grieser, Repnow, Schnell, Schwalm,
  Wolf, Zhou, Kieslich, M\"{u}ller, Schippers, Colgan, Loch, Chen, and
  Gu}]{Savin2006a}
Savin, D.~W., Gwinner, G., Grieser, M., Repnow, R., Schnell, M., Schwalm, D.,
  Wolf, A., Zhou, S.-G., Kieslich, S., M\"{u}ller, A., Schippers, S., Colgan,
  J., Loch, S.~D., Chen, M.~H., Gu, M.~F., 2006. Dielectronic recombination of
  {F}e {XXXIII} forming {F}e {XXII}: {L}aboratory measurements and theoretical
  calculations. Astrophys. J. 642, 1275.

\bibitem[{Schippers(2009{\natexlab{a}})}]{Schippers2009a}
Schippers, S., 2009{\natexlab{a}}. Astrophysical relevance of storage-ring
  electron-ion recombination experiments. J. Phys.: Conf. Ser. 163, 012001.

\bibitem[{Schippers(2009{\natexlab{b}})}]{Schippers2009}
Schippers, S., 2009{\natexlab{b}}. Relativistic, {QED} and nuclear effects in
  highly charged ions revealed by resonant electron-ion recombination in
  storage rings. Nucl. Instrum. Methods B 267, 192.

\bibitem[{Schippers(2012)}]{Schippers2012a}
Schippers, S., 2012. Storage-ring ionization and recombination experiments with
  multiply charged ions relevant to astrophysical and fusion plasmas. J. Phys.:
  Conf. Ser. 388, 012010.

\bibitem[{Schippers(2013)}]{Schippers2013}
Schippers, S., 2013. Storage-ring measurements of hyperfine induced transition
  rates in berylliumlike ions. AIP Conf. Proc. 1545, 7.

\bibitem[{Schippers et~al.(2011)Schippers, Bernhardt, M\"{u}ller, Krantz,
  Grieser, Repnow, Wolf, Lestinsky, Hahn, Novotn\'{y}, and
  Savin}]{Schippers2011}
Schippers, S., Bernhardt, D., M\"{u}ller, A., Krantz, C., Grieser, M., Repnow,
  R., Wolf, A., Lestinsky, M., Hahn, M., Novotn\'{y}, O., Savin, D.~W., 2011.
  Dielectronic recombination of xenonlike tungsten ions. Phys. Rev. A 83,
  012711.

\bibitem[{Schippers et~al.(2012)Schippers, Bernhardt, M\"{u}ller, Lestinsky,
  Hahn, Novotn{\' y}, Savin, Grieser, Krantz, Repnow, and Wolf}]{Schippers2012}
Schippers, S., Bernhardt, D., M\"{u}ller, A., Lestinsky, M., Hahn, M.,
  Novotn{\' y}, O., Savin, D.~W., Grieser, M., Krantz, C., Repnow, R., Wolf,
  A., 2012. Storage-ring measurement of the hyperfine-induced $2s\,2p\;^3{P}_0
  \to 2s^2\;^1{S}_0$ transition rate in berylliumlike sulfur. Phys. Rev. A 85,
  012513.

\bibitem[{Schippers et~al.(2010)Schippers, Lestinsky, M\"{u}ller, Savin,
  Schmidt, and Wolf}]{Schippers2010}
Schippers, S., Lestinsky, M., M\"{u}ller, A., Savin, D.~W., Schmidt, E.~W.,
  Wolf, A., 2010. Dielectronic recombination data for astrophysical
  applications: {P}lasma rate-coefficients for {F}e$^{q+}$ ($q$=7–10,
  13–22) and {N}i$^{25+}$ ions from storage-ring experiments. Int. Rev. At.
  Mol. Phys. 1, 109.

\bibitem[{Schippers et~al.(2001)Schippers, M\"{u}ller, Gwinner, Linkemann,
  Saghiri, and Wolf}]{Schippers2001c}
Schippers, S., M\"{u}ller, A., Gwinner, G., Linkemann, J., Saghiri, A.~A.,
  Wolf, A., 2001. Storage ring measurement of the {C\,IV} recombination rate
  coefficient. Astrophys. J. 555, 1027.

\bibitem[{Schippers et~al.(2007)Schippers, Schmidt, Bernhardt, Yu, M\"{u}ller,
  Lestinsky, Orlov, Grieser, Repnow, and Wolf}]{Schippers2007a}
Schippers, S., Schmidt, E.~W., Bernhardt, D., Yu, D., M\"{u}ller, A.,
  Lestinsky, M., Orlov, D.~A., Grieser, M., Repnow, R., Wolf, A., 2007.
  Storage-ring measurement of the hyperfine induced
  $^{47}${T}i$^{18+}$(${2s\,2p\;^3P_0 \to 2s^2\;^1S_0}$) transition rate. Phys.
  Rev. Lett. 98, 033001.

\bibitem[{Schippers et~al.(2004)Schippers, Schnell, Brandau, Kieslich,
  M\"{u}ller, and Wolf}]{Schippers2004c}
Schippers, S., Schnell, M., Brandau, C., Kieslich, S., M\"{u}ller, A., Wolf,
  A., 2004. Experimental {M}g {IX} photorecombination rate coefficient. Astron.
  Astrophys. 421, 1185.

\bibitem[{Schnell et~al.(2003)Schnell, Gwinner, Badnell, Bannister, B\"{o}hm,
  Colgan, Kieslich, Loch, Mitnik, M\"{u}ller, Pindzola, Schippers, Schwalm,
  Shi, Wolf, and Zhou}]{Schnell2003b}
Schnell, M., Gwinner, G., Badnell, N.~R., Bannister, M.~E., B\"{o}hm, S.,
  Colgan, J., Kieslich, S., Loch, S.~D., Mitnik, D., M\"{u}ller, A., Pindzola,
  M.~S., Schippers, S., Schwalm, D., Shi, W., Wolf, A., Zhou, S.-G., 2003.
  Observation of trielectronic recombination in {B}e-like {C}l ions. Phys. Rev.
  Lett. 91, 043001.

\bibitem[{Schuch and B\"{o}hm(2007)}]{Schuch2007a}
Schuch, R., B\"{o}hm, S., 2007. Atomic physics with ions stored in the round.
  J. Phys.: Conf. Ser. 88, 012002.

\bibitem[{Schwalm(2007)}]{Schwalm2007}
Schwalm, D., 2007. Atomic and molecular astrophysics with heavy ion storage
  rings. Progress in Particle and Nuclear Physics 59, 156.

\bibitem[{Scully et~al.(2005)Scully, Emmons, Gharaibeh, Phaneuf, Kilcoyne,
  Schlachter, Schippers, M\"{u}ller, Chakraborty, Madjet, and
  Rost}]{Scully2005a}
Scully, S. W.~J., Emmons, E.~D., Gharaibeh, M.~F., Phaneuf, R.~A., Kilcoyne, A.
  L.~D., Schlachter, A.~S., Schippers, S., M\"{u}ller, A., Chakraborty, H.~S.,
  Madjet, M.~E., Rost, J.~M., 2005. Photoexcitation of a volume plasmon in
  {C}$_{60}$ ions. Phys. Rev. Lett. 94, 065503.

\bibitem[{Spruck et~al.(2014)Spruck, Badnell, Krantz, Novotn\'{y}, Becker,
  Bernhardt, Grieser, Hahn, Repnow, Savin, Wolf, M\"{u}ller, and
  Schippers}]{Spruck2014a}
Spruck, K., Badnell, N.~R., Krantz, C., Novotn\'{y}, O., Becker, A., Bernhardt,
  D., Grieser, M., Hahn, M., Repnow, R., Savin, D.~W., Wolf, A., M\"{u}ller,
  A., Schippers, S., 2014. Recombination of {W}$^{18+}$ ions with electrons:
  {A}bsolute rate coefficients from a storage-ring experiment and from
  theoretical calculations. Phys. Rev. A, submitted.

\bibitem[{{St\"{o}hlker} et~al.(2014){St\"{o}hlker}, {Litvinov},
  {Br\"{a}uning-Demian}, {Lestinsky}, {Herfurth}, {Maier}, {Prasuhn}, {Schuch},
  {Steck}, and {for the SPARC Collaboration}}]{Stoehlker2014}
{St\"{o}hlker}, T., {Litvinov}, Y.~A., {Br\"{a}uning-Demian}, A., {Lestinsky},
  M., {Herfurth}, F., {Maier}, R., {Prasuhn}, D., {Schuch}, R., {Steck}, M.,
  {for the SPARC Collaboration}, 2014. {SPARC Collaboration: New Strategy
  for Storage Ring Physics at FAIR}. Hyperfine Interact. 227, 45.

\bibitem[{West and Morton(1978)}]{West1978}
West, J.~B., Morton, J., 1978. Absolute photoionization cross-section tables
  for xenon in the {VUV} and the soft x-ray regions. At. Data Nucl. Data Tables
  22, 103.

\bibitem[{Wolf et~al.(2006)Wolf, Buhr, Grieser, {von Hahn}, Lestinsky,
  Lindroth, Orlov, Schippers, and Schneider}]{Wolf2006c}
Wolf, A., Buhr, H., Grieser, M., {von Hahn}, R., Lestinsky, M., Lindroth, E.,
  Orlov, D.~A., Schippers, S., Schneider, I.~F., 2006. Progress in stored ion
  beam experiments on atomic and molecular processes. Hyperfine Interact. 172,
  111.

\end{thebibliography}

\end{document}